\documentclass[aps,prc,showpacs,twocolumn,preprintnumbers,floatfix, showkeys,tightenlines]{revtex4}
\usepackage{amsmath,amssymb,amsfonts}
\usepackage{hyperref}
\usepackage{epsfig,subfigure,latexsym}
\usepackage{rotating}
\usepackage{graphicx}
\usepackage{amsmath}
\usepackage{color}

\allowdisplaybreaks

\begin{document}

\title{Reassessing nuclear matter incompressibility and its density dependence}

\author{J. N. De}
\address{Saha Institute of Nuclear Physics, 1/AF Bidhannagar, Kolkata
{\sl 700064}, India}
\author{S. K. Samaddar}
\address{Saha Institute of Nuclear Physics, 1/AF Bidhannagar, Kolkata
{\sl 700064}, India}
\author{B. K. Agrawal}
\address{Saha Institute of Nuclear Physics, 1/AF Bidhannagar, Kolkata
{\sl 700064}, India}

\begin{abstract}
Experimental giant monopole resonance energies are now known to constrain
nuclear incompressibility of symmetric nuclear matter $K$ and its
density slope $M$ at a particular value of sub-saturation density, the
crossing density $\rho_c$. Consistent with these constraints, we propose
a reasonable way to construct a plausible equation of state of symmetric
nuclear matter in a broad density region around the saturation density
$\rho_0$. Help of two additional empirical inputs, the value of $\rho_0$
and that of the energy per nucleon $e(\rho_0)$ are needed. The
value of $K(\rho_0)$ comes out to be $211.9\pm 24.5$ MeV.
\end{abstract}
\pacs {21.10.Re,21.65.-f,21.65.Mn}
\maketitle

\section{Introduction}

The nuclear incompressibility parameter $K_0$ defined for symmetric
nuclear matter (SNM) at saturation density $\rho_0 $ stands out as
an irreducible element of physical reality. It  has an umbilical association with
the isoscalar giant monopole resonances (ISGMR) in
microscopic nuclei; it also underlies in a proper understanding of
supernova explosion in the cosmic domain \cite{Brown88}. From careful
microscopic analysis of ISGMR energies with suitably constructed energy density
functional (EDF) ${\cal E}(\rho) $ in a non relativistic framework as applicable to finite and infinite
nuclear systems, its value had initially been fixed at $K_0 \simeq 210 \pm
30 $MeV \cite{Blaizot80,Farine97}.  In microscopic relativistic approaches
on the other hand, a higher value of $K_0$ $\sim 260$ MeV was obtained
\cite{Vretenar03}. After several revisions from different corners, however,  its value settled
to $K_0 \simeq 230 \pm 20 $ MeV \cite{Todd-Rutel05,Avogadro13,Niksic08}.
It gives good agreement with the experimentally determined centroids of
ISGMR, in particular, for $^{208}$Pb , $^{90}$Zr and $^{144}$Sm nuclei,
calculated both with non-relativistic \cite{Agrawal05,kvasil14} and
relativistic \cite{Niksic08} energy density functionals. The near-settled
problem was, however, left open  with the apparent incompatibility of
the said value of $K_0$ with the recent ISGMR data for Sn and Cd-isotopes
\cite{Lui04,Youngblood04,Li07,Piekarewicz07,Garg07,Tselyaev09,Li10,Patel12,Avogadro13}.
These nuclei showed remarkable softness towards compression, the ISGMR
data appeared  explained best  with $K_0 \sim $200 MeV \cite{Avogadro13}.

A plausible explanation was recently put forward by Khan $\it{et. al }$
\cite{Khan12} for the apparent discrepancy. It is argued that there may
not be an unique relation between the value of	$K_0$ associated with an
effective force and the monopole energy of a nucleus predicted by the
force \cite{Colo04a}.  The region between the center and the surface of
the nucleus is the most sensitive towards displaying the compression
as manifested in the ISGMR. The ISGMR centroid $E_G$ is related to
the integral of incompressibility $(\int K(\rho )d\rho )$ over the
whole density range \cite{Khan10}.  As a result, a larger value
of  $K(\rho_0)$ for a given EDF can be compensated by lower values
of $K(\rho)$ at sub-saturation densities so as to predict a similar
value of ISGMR energy in nuclei. It is seen that the incompressibility
$K(\rho )$ calculated with a multitude of energy density functionals
when plotted against density cross close to a single density point \cite{Khan12},
this  universality possibly arising from the constraints encoded in the
EDF from empirical nuclear observables.  This crossing density $\rho_c
[= (0.71 \pm 0.005) \rho_0] $ \cite{Khan13} seems more relevant as an indicator for the ISGMR
centroid. Because of the incompressibility integral, the centroid seems
more intimately correlated to the derivative of the compression modulus
(defined as $M=3\rho K^\prime(\rho )$ ) at the crossing density rather
than to $K_0$. The value of $K_c (=K(\rho_c))$ is seen to be $\sim 35 \pm 4$
MeV \cite{Khan13}.  From various functionals, the calculated values
of $M_c(=M(\rho_c))$ are found to be linearly correlated with the
correspondingly calculated values of ISGMR centroids for $^{208}$Pb
and also for  $^{120}$Sn. From the known experimental ISGMR data for
these nuclei, a value of $M_c \simeq $1050 $\pm $100 MeV \cite{Khan13}
is then obtained, revised from an earlier estimate of $ 1100 \pm 70 $
MeV \cite{Khan12}.  Using a further assumption of a  linear correlation
between $K_0$ and $E_G$ calculated from different EDF,	a value  for $K_0$
$\simeq $230 MeV with an uncertainty of $\simeq $40 MeV is reported,
the uncertainty being inferred from the spread of $K_0$ values obtained
with the different functionals used.

The universality of the crossing point $\rho_c$ and the values of $K_c$
and $M_c$ can be readily acknowledged; $M_c$ is seen to be well correlated
to $E_G$. The Pearson correlation coefficient $r$ \cite{Brandt97} of $M_c$ with $E_G$
for  $^{120}$Sn is 0.80 and is  0.94 for $^{208}$Pb.
However, assumption of a linear correlation between $K_0$ and $E_G$
may not be justified, they seem to be very weakly correlated ($r$=0.67
for $^{120}$Sn and  0.79 for $^{208}$Pb) \cite{Khan13}. The inferred
value of incompressibility around saturation may then be called into
question.  One can see that a  linear Taylor expansion $K_0(\rho_0)
=K(\rho_c)+(\rho_0-\rho_c)K^\prime ( \rho_c) $ yields for $K_0 \simeq
185\pm 14.3 $ MeV, noting that $K^\prime (\rho_c)=M_c/(3\rho_c)$.

The absence of a strong linear correlation between $K_0$ and $E_G$
calculated from different effective forces prompts one to think that $K_c$
and $M_c$ alone are not sufficient to yield the correct value of $K_0$.
Further empirical  information is possibly needed to arrive at that.
In this paper, we show that with given values of only $K_c$
and $M_c$ along with some time-tested values of empirical nuclear
constants, it is possible to address to a proper assessment of the value
of incompressibility $K$ and its density dependence.  The empirical
constants are  the saturation density $\rho_0$, taken as 0.155 $\pm $
0.008 fm$^{-3}$ for SNM and the energy per nucleon at that density
$e(\rho_0)$,  taken as $-16.0 \pm $0.1 MeV \cite{Agrawal12, Alam14}.
An acceptable value of the effective nucleon mass $m^*/m $, which lies
in the range $m^*/m \sim $0.8 $\pm $0.2 \cite{Dutra12} at saturation
density is also used.

This paper is structured as follows. In Sec. II, we introduce the theoretical elements to calculate 
the nuclear equation of state from $K_c$ and $\rho_c$ with the aid of empirical inputs mentioned.
Results and discussions are presented in Sec. III. Sec. IV contains the concluding remarks.

\section{Theoretical Edifice}
	We keep the discussions pertinent for SNM 
        at any
        density $\rho $ at zero temperature ($T=0 $). 
        The chemical potential of a nucleon is
	given by the single-particle energy at the Fermi surface,
	\begin{eqnarray}
	\mu=\varepsilon_F=\frac{p_F^2}{2m}+U
	\end{eqnarray}
	where $p_F(\rho)$ is the Fermi momentum and $U(\rho)$  the single-particle
	potential. Assuming the nucleonic interaction to be momentum
	and density dependent, the single-particle potential separates
	into three parts \cite{Bandyopadhyay90}
	\begin{eqnarray}
	U=V_0+p_F^2V_1+V_2 .
	\end{eqnarray}
	The last term $V_2$ is the rearrangement potential that arises
	only for density-dependent interactions, and the second is the
	momentum-dependent term that defines the effective mass $m^*$,
	\begin{eqnarray}
	\frac{p_F^2}{2m^*}=\frac{p_F^2}{2m}+p_F^2V_1
	\end{eqnarray}
	so that
	\begin{eqnarray}
	\frac{1}{m^*}=\frac{1}{m}+2V_1.
	\end{eqnarray}
	The energy per nucleon at density $\rho $ is given by,
	\begin{eqnarray}
	e=<\frac{p^2}{2m}>+\frac{1}{2}<p^2>V_1+\frac{1}{2}V_0  \nonumber \\
	=\frac{1}{2}(1+\frac{m^*}{m})<\frac{p^2}{2m^*}>+\frac{1}{2}V_0
	\end{eqnarray}
	From Gibbs-Duhem relation,
	\begin{eqnarray}
	\mu=e+\frac{P}{\rho },
	\end{eqnarray}
	where $P$ is the pressure. Keeping this in mind, from Eqs.~(1),(5) and
	(6), we get
	\begin{eqnarray}
	e(\rho )=\frac{p_F^2}{10m}[3-\frac{2m}{m^*}]-V_2
	+\frac{P}{\rho},                  
	\end{eqnarray}
	where we have put $<p^2>=\frac{3}{5}p_F^2$.

	The density dependence of the effective 
        mass \cite{Bohr69} can be cast as
	$\frac{m}{m^*}=1+k\rho $, the rearrangement potential can be written
	in the form $V_2=a\rho^\alpha $. This is the
        form that emerges for finite range density-dependent 
        forces \cite{Bandyopadhyay90} in a non relativistic framework
        or for Skyrme interactions.
	The quantities $a,~\alpha $ and
	$k$ are numbers. If $\frac{m^*}{m} (\rho_0)$ is chosen, $k$
	is known.

	At $\rho =\rho_0 , P=0 $, then from Eq.~(7), 
        writing for  $\frac{p_F^2}{2m}
	=b\rho^{2/3}$ with
	$b=\frac{(\frac{3}{2}\pi^2)^{2/3}\hbar^2} {2m}$,
	\begin{eqnarray}
	e_0=e(\rho_0)=\frac{b}{5}\rho_0^{2/3}[1-2k\rho_0]-a\rho_0^{\alpha}.
	\end{eqnarray}
	Since $P=\rho^2\frac{\partial e}{\partial \rho }$, from Eq.~(7) again
	we get,
	\begin{eqnarray}
	P=\frac{b}{15}\rho^{5/3}-\frac{1}{3}bk\rho^{8/3}-\frac{1}{2}
	\alpha a \rho^{\alpha +1}+
	\frac{1}{2}\rho \frac{\partial P}{\partial \rho}.
	\end{eqnarray}
	At $\rho_0$, this yields (since $K_0=9 \frac{\partial P}{\partial \rho}|_{\rho_0}$),
	\begin{eqnarray}
	\frac{1}{2} \alpha a\rho_0^\alpha +\frac{1}{3}bk\rho_0^{5/3}
	-(\frac{K_0}{18}+\frac{b}{15}\rho_0^{2/3})=0.
	\end{eqnarray}
	Furthermore, Eq.~(9) gives
	\begin{eqnarray}
	K (\rho )&=&9\frac{\partial P}{\partial \rho }=2b\rho^{2/3}
	-16bk\rho^{5/3}\nonumber \\
        &&-9\alpha (\alpha +1)a\rho^\alpha +9\rho \frac{\partial^2 P}
	{\partial \rho^2}.
	\end{eqnarray}
	Defining $M=3\rho \frac{dK}{d\rho }=27 \rho \frac{\partial^2 P}{
	\partial \rho^2}$, this leads, at $\rho =\rho_c $ to
	\begin{eqnarray}
	9\alpha (\alpha +1)a\rho_c^\alpha +16bk\rho_c^{5/3}-(2b\rho_c^{2/3}
	+\frac{M_c}{3}-K_c)=0.
	\end{eqnarray}
	Since $k$ is a given entity and $\rho_c$ and $(M_c/3-K_c )$ are known,
	eqs.~(8) and (12) can be solved for $a$ and $\alpha $, eq.~(10)  then
	gives the value of the nuclear incompressibility $K_0$. Once $K_0$ is
	obtained, $M_0 (=M(\rho_0))$ is evaluated from eq.~(12) by choosing
	$\rho_0$ for $\rho_c$. Then
	$Q_0=27 \rho_0^3\frac {\partial ^3 e}{\partial \rho^3}|_{\rho_0}$ is
	also known from $M_0=12K_0+Q_0$.

The structure of eq.~(9) shows that the pressure and its first
derivative are interrelated.
One can then get higher density derivatives of $P$ or of energy 
$e$ recursively  from
eq.~(9) as is evident from eq.~(11). For the present, we show that 
\begin{eqnarray}
9\rho \frac{\partial ^3 P}{\partial \rho^3}=
9\alpha^2(\alpha +1)a\rho^{\alpha -1}+\frac{80}{3}bk\rho^{2/3}
-\frac{4}{3}b\rho^{-1/3}.
\end{eqnarray}
Since 
\begin{eqnarray}
\frac{\partial ^3P}{\partial \rho^3}=6\frac{\partial ^2e}{\partial \rho^2}
+6\rho \frac{\partial ^3e}{\partial \rho^3} +\rho^2\frac{\partial ^4e}
{\partial \rho^4},
\end{eqnarray}
we find
\begin{eqnarray}
9\rho_0^2 \frac{\partial ^3 P}{\partial \rho^3}|_{\rho_0}=
6K_0+2Q_0 +\frac{1}{9}N_0.
\end{eqnarray}
where we have defined $N_0=81 \rho_0^4 \frac{\partial^4 e}{\partial
\rho^4}|_{\rho_0}$. From eq.~(13) and (15), knowing $K_0$ and $Q_0$,
$N_0$ can be calculated.
Similarly, one can calculate the fifth density derivative
of energy ($R_0=243 \rho_0^5\frac{\partial^5 e}{\partial \rho^5}|_{\rho_0}$)
by exploiting eqs.~(13) and (14) from
\begin{eqnarray}
9\rho_0^3\frac{\partial^4 P}{\partial \rho^4}|_{\rho_0}=4Q_0+
\frac{8}{9}N_0+\frac{1}{27}R_0.
\end{eqnarray}
These help to find the density variation of the
energy and also of the incompressibility, as is seen,
\begin{eqnarray}
e(\rho)&=&e(\rho_0)+\frac{1}{2}K_0\epsilon^2+\frac{1}{6}Q_0
\epsilon^3\nonumber \\
&&+\frac{1}{24}N_0 \epsilon^4 + \frac{1}{120}R_0
\epsilon^5 + ...~~~,
\end{eqnarray}
 where $\epsilon =(\frac{\rho-\rho_0}{3\rho_0})$  
(counting terms only up to $\epsilon^5 $ is seen 
to be a very good approximation in the  density  range of $\sim $
$\rho_0/4 <\rho <2.0 \rho_0$, we retain terms up to them). 
Eqs.~ (7) and (17)  give
\begin{eqnarray}
\frac{P(\rho)}{\rho}&=&e(\rho_0)+\frac{1}{2}K_0\epsilon^2+
\frac{1}{6}Q_0\epsilon^3
+\frac{1}{24}N_0\epsilon^4\nonumber \\
&&+\frac{1}{120}R_0\epsilon^5
-\frac{b}{5}\rho^{2/3}[1-2k\rho]+a\rho^\alpha.
\end{eqnarray}
and eq.~(9) gives
\begin{eqnarray}
K(\rho )=9\frac{dP}{d\rho }=18[\frac{P}{\rho }-\frac{b}{15}\rho^{2/3}
+\frac{1}{3}bk\rho^{5/3}+\frac{1}{2}\alpha a\rho^\alpha ].
\end{eqnarray}
We have thus the equation of state (EOS) of symmetric nuclear 
matter in a reasonably spread-out density domain around the 
saturation density.

\begin{figure}[h]
\centerline{\includegraphics[width=0.9\linewidth]{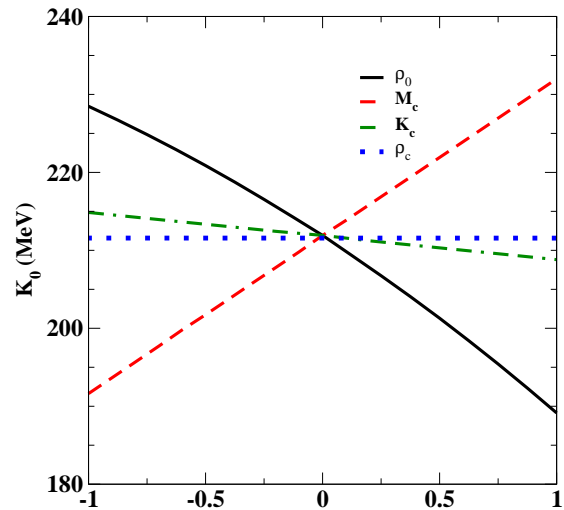}}
\caption{(Color online)
\label{fig1}
The sensitivity of the incompressibility $K_0$ at the saturation density
$(\rho_0)$ on the values of the incompressibility $K_c$ (green dash-dotted
line), its density slope $M_c$ (red dashed line), the crossing density
$\rho_c$ (blue dotted line) and the value of $\rho_0$ (black full line).
The abscissa extends from -1 to +1.  These end points refer to the
scaled lower and upper limits of $K_c$, $M_c$, $\rho_c$ and $\rho_0$,
respectively (see text).}
 \end{figure}

The incompressibility $K$ at any density $\rho$ can be
calculated directly from eq.~(19) or it may be calculated
in terms of $K(\rho_c)$ and its higher density derivatives as
\begin{eqnarray} K(\rho)&=&K(\rho_c)+(\rho-\rho_c)K^\prime(\rho_c)
+\frac{(\rho-\rho_c)^2}{2}K^{\prime \prime}(\rho_c)\nonumber \\ &&
+\frac{(\rho-\rho_c)^3}{6}K^{\prime \prime \prime }(\rho_c)+ ... .
\end{eqnarray} The different derivatives can be calculated from eq.~(19).
With given  values of $ \rho_0, e_0, \frac{m^*}{m}(\rho_0),$ and $\rho_c$,
one notes that the solutions for $a$ and $\alpha$ do not depend separately
on $K_c$ and $ M_c$, but on $(M_c/3-K_c)$.

\section{Results and Discussions}

The values of the empirical constants $\rho_0$, $e_0$ and $\frac{m^*}{m}$ needed for our calculation
have already been mentioned. As for the crossing density, 
we choose $\rho_c=0.110 \pm 0.0008$ fm$^{-3}$. 
With given inputs of $M_c$ and $K_c$, it should be noted  that the
output values for $M_c$ and $K_c$ may come out to be different, but
$(M_c/3-K_c)$ remains invariant. With inputs $M_c$=1050 MeV and $K_c$
=35 MeV, the output $M_c$ and $K_c$ are found to be 1051.8 MeV and 35.46
MeV, respectively. Since they are very close to the input values, they
were not tinkered with for exact matching of the output and input values.
The value of incompressibility at $\rho_0$ turns out to be $K_0$=211.9$\pm
$24.5 MeV either from eq.~(19) or eq.~(20). We note that in eq.~(20),
at saturation, the value of the second term on the right hand side is
143.3 MeV, the third term is 35.9 MeV, the fourth term is $-$3.2 MeV,
the fifth term (not shown in eq.~(20)) is 0.55 MeV and so on,which adds
up to $\sim $ 211.9 MeV.

The uncertainty in an observable $X$ (like
$K, M$ etc) is calculated from $\Delta X^2=\sum_i 
(\frac{\partial X}{\partial y_i}\Delta y_i)^2 $ where
$\Delta y_i$ are the uncertainties in the empirically
known entities $y_i$. The sensitivity of $K_0$ on these entities that
influence the incompressibility most is displayed in Fig.\ref{fig1}.
The abscissa is scaled such that $0$ refers to the central value of
these entities $M_c, K_c, \rho_c$  and $\rho_0$; $\pm 1$ refer to the
extrema of their domain ($\pm 100$ MeV, $\pm 5$ MeV, $\pm 0.005 \rho_0$
and $\pm 0.008$ fm$^{-3}$ from the central values  of the entities,
respectively). The value of $K_0$ is seen to be very sensitive with
changes in either $M_c$ or $\rho_0$ when all other input entities are
kept fixed. Its sensitivity to $K_c$ or $\rho_c$ is weak; on $\frac
{m^*}{m}$ or to the energy per nucleon $e_0$, it is rather insensitive.
The near-insensitivity of incompressibility to the effective mass is
observed for Skyrme density functionals also. From the data base for
these functionals as tabulated by Dutra {\it {et. al}} \cite{Dutra12},
the correlation coefficient between $K_0$ and $m^*$ is calculated to be
only $\sim -$0.2.

\begin{figure}[h]
\centerline{\includegraphics[width=0.9\linewidth]{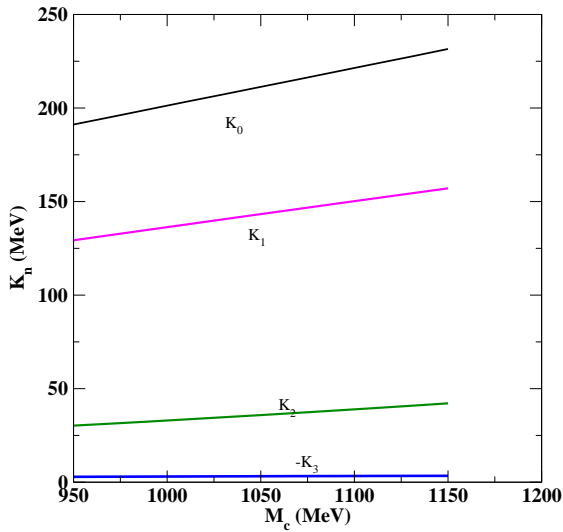}}
\caption{(Color online)}
\label{fig2}
The  incompressibility and its different density derivative as defined in the text plotted as a
function of $M_c$.
\end{figure} 

The near-perfect linear correlation of $K_0$ with $M_c$ as seen in
Fig.\ref{fig1} is very startling. From Eq.~(20), one may expect that the
second and higher order derivatives of $K(\rho_c)$ would destroy
this correlation. However, we find that both $K^{\prime \prime
}$ and $K^{\prime \prime \prime }$ are also linearly correlated
with $M_c$ and thus $K(\rho_0)$ retains its linear correlation
with $M_c$. This is displayed in Fig.\ref{fig2}, where we define $K_1=
(\rho_0-\rho_c)K^\prime(\rho_c)$, $K_2=\frac{(\rho_0-\rho_c)^2}{2}K^{\prime
\prime}(\rho_c)$ and $K_3= \frac{(\rho_0-\rho_c)^3}{6}K^{\prime \prime
\prime }(\rho_c)$.  The weak correlation between $K_0$ and $M_c$ that
can be inferred from the calculated correlation structure of $(M_c-E_G)$
and $(K_0-E_G)$ in refs.\cite{Khan12,Khan13} possibly results from the
use of different EDFs in getting the various relevant observables.

\begin{figure}[h]
\centerline{\includegraphics[width=0.9\linewidth]{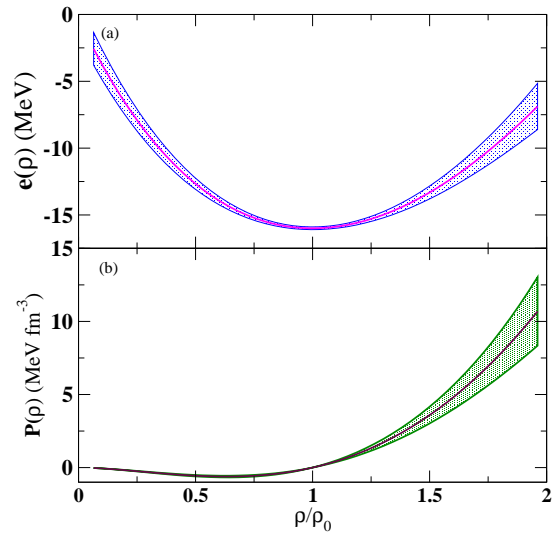}}
\caption{(Color online)
\label{fig3}
The nuclear EOS as a function of density. The panels (a) and  (b) 
show the energy per nucleon and  pressure, 
respectively in a selected range around the saturation density.}
\end{figure} 
\begin{figure}[h]
\centerline{\includegraphics[width=0.9\linewidth]{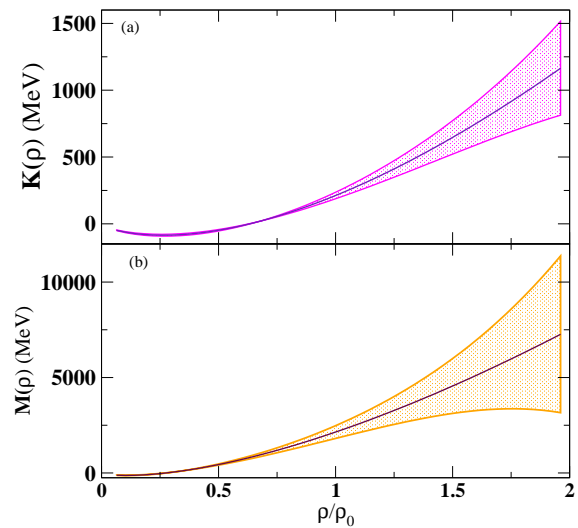}}
\caption{(Color online)
\label{fig4}
The nuclear EOS as a function of density. The panels (a) and (b)
show the  incompressibility and its density derivative $M$,
respectively in a selected range around the saturation density.}
\end{figure} 
Figures \ref{fig3} and \ref{fig4}  display the functional dependence of
the nuclear EOS on density. The panels (a) and (b) in Fig.\ref{fig3} show the
energy per nucleon and the pressure, the ones in Fig.\ref{fig4} show the
incompressibility and its density derivative $M$, respectively. As one
sees, the uncertainty in energy and pressure  grows as one moves away from
the saturation density, similarly the uncertainty in incompressibility or
its density derivative increases with distance from the crossing density.

\section{Conclusions} 

To sum up, we have made a  modest attempt to reassess the value
of $K(\rho_0)$ consistent with the new-found constraint on the
incompressibility  $K(\rho_c)$ and its density slope $M(\rho_c)$ at a
particular value of density at sub-saturation, the crossing density
$\rho_c$. We have relied   on  some empirically well-known values
of nuclear constants.  We have further made the assumption of linear
density dependence of the effective mass and the power law dependence
of the rearrangement potential which happens to be generally  true
for non-relativistic momentum and density dependent interactions.
In relativistic models, the density dependence of the effective mass
may not be linear \cite{Serot86}.  The rearrangement potential appears
explicitly there only in the case of density dependent meson exchange
models \cite{Typel99}.

The value of incompressibility $K(\rho_0)$ turns out to be $211.9 \pm
24.5$ MeV. This is somewhat lower than  the current value in vogue, $K_0
\sim 230 \pm 20$ MeV.  From recursive relations, our method  allows also
estimates of higher density derivatives of energy or of  pressure and
thus helps in constructing the nuclear EoS $e(\rho)$ at and around the
saturation density.

The authors gratefully acknowledge the assistance of Tanuja 
Agrawal in the preparation of the manuscript.
One of the authors (JND) acknowledges support from the Department
of Science $\& $ Technology, Government of India.



\end{document}